\documentstyle[aps,pre,twocolumn,psfig]{revtex}   
\begin{document}

\def\seq{\vspace{0cm}}

%\hsize\textwidth\columnwidth\hsize\csname @twocolumnfalse\endcsname
\title{Recent results on Multiplicative Noise}
\author{Walter Genovese$^{1}$ and Miguel A. Mu{\~{n}}oz $^{1,2}$}
\address{ $^{1}$ I.N.F.M. Sezione di Roma and Dipartimento di Fisica,
Universit\'a di Roma ``La Sapienza'',~P.le
A. Moro 2, I-00185 Roma, Italy}
\address{ $^{2}$ The Abdus Salam International
Centre for Theoretical Physics (ICTP)
P.O. Box 586, 34100 Trieste, Italy}
\date{\today }
\maketitle

\begin{abstract}
   
Recent developments in the analysis of 
Langevin equations with multiplicative
noise (MN) are reported.
In particular, we:
 (i) present
 numerical simulations in three dimensions showing
that the MN equation exhibits,
 like the Kardar-Parisi-Zhang (KPZ) equation both a weak
coupling fixed point and a strong coupling phase, supporting
the proposed
relation between MN and KPZ;
 (ii) present  
 dimensional,  
 and mean field analysis of the MN equation to compute critical
exponents;
 (iii) show that the phenomenon of the noise induced 
ordering transition associated with the MN equation appears
only in the Stratonovich representation and not in the
Ito one, and
 (iv) report the presence of a new first-order like 
phase transition at zero spatial coupling, supporting the fact that 
this is the minimum model for noise induced ordering transitions.
\vspace{8pt}
\vspace{8pt}

\end{abstract}

\begin{text}
\narrowtext

\section{Introduction}

The idea that
noise can induce rather non-trivial 
 effects when added to
deterministic equations
it is not considered any more a shocking one. 
 Some recently undercovered phenomena have
familiarized us with the idea that strange physical 
mechanisms induced by noise are not as infrequent previously
thought.
Stochastic resonance \cite{sr},
resonant activation \cite{sa}, 
noise-induced spatial patterns \cite{n1}, 
noise-enhanced multistability in coupled oscillators
 \cite{Park}, and noise-induced 
phase transitions \cite{Raul,BK,joan,GMG} are just a few examples.
 In particular, a lot of attention has been devoted in recent years
 to the
study of phase transitions appearing in systems the associated
deterministic part of which does not exhibit any symmetry breaking.
 These studies
were mostly limited to one-variable systems  \cite{HL} 
until an interesting paper by Van den Broeck,
 Parrondo and Toral ) \cite{Raul,Kawai} 
(see also \cite{BK}).  These authors showed the possibility of having 
noise-induced transitions in spatially extended systems, and 
illustrated the physical mechanism 
originating this phenomenon: A short
time instability is generated owing to the noise,
and the generated non-trivial state 
 is afterwards rendered 
stable by the spatial coupling \cite{Kawai}.
 In this way, by
increasing the noise amplitude the instability in enhanced, and
the system becomes
more and more ordered:
A noise-induced ordering phase transition (NIOT) is generated.
In the model presented in  \cite{Raul} the NIOT was followed on
further increasing of the noise amplitude by a           
 second phase transition. 
At larger noise amplitudes,
the usual role of the noise
as a disorganizing source takes over and the system becomes again
disordered.
 This is what we call a 
 noise-induced disordering transition (NIDT).
The same type of behavior
 has been found in other models
\cite{Sancho,Kim,Muller}.

   In a recent paper \cite{GMS} (see also \cite{Com}) we put forward that 
the NIOT and the NIDT have different origin. The NIOT is induced by
multiplicative noise, while the NIDT is due to the presence
of additive noise 
(even though it can also be generated in a somehow 
artificial way by multiplicative noise \cite{GMS}).
 In this way 
we proposed  the Langevin
equation with multiplicative noise \cite{MN,MN2},
interpreted in the Stratonovich 
sense, as a 
 new minimal model for NIOT.
 As a consequence of the previous 
observation we predicted and afterward confirmed the existence
 NIOTs in one-dimensional systems.

 On the other
hand, the MN Langevin equation has been proved to be related 
to the Kardar-Parisi-Zhang (KPZ) equation describing non-equilibrium
surface growth \cite{KPZ}.  In fact, by performing a, so called,
Cole-Hopf transformation the MN Langevin equation
becomes the KPZ equation with an extra wall that limits the 
maximum value of the height \cite{MN,MN2,DNA}. In this way,
the critical point of the MN equation is related to a
{\it wetting transition}.
In fact, for large values of the control parameter,
the surface escapes from the limiting wall and behaves as a KPZ
surface, while for smaller values of the control parameter
 there is a phase in which the surface
 {\it wets} the wall and remains bound to it.
 Separating
both phases there is a critical point at which the surface gets
depinned or unbound \cite{MN2,Firenze}.
 This critical point may either be a
weak or a strong coupling fixed point depending on 
the noise intensity and on the system dimensionality.

   In this paper we continue to explore  the Langevin equation with 
 multiplicative noise from different perspectives. 
The paper is structured as follows.

 In section II we
present the MN Langevin equation, discuss
 its connection
with KPZ,  and define the critical exponents.

In section III, we present
dimensional analysis and predictions for the mean field exponents.

In section IV,
by exploiting
the connection with KPZ we try to observe numerically
whether
 the MN equation does exhibit strong noise
 and weak noise fixed points 
\cite{KPZ,Terry,HZ,Laszlo} in 
dimensions larger than two.

In section V we analyze the MN equation from the  Ito-Stratonovich 
dilemma point of view and find out that the NIOT is specific
of the Stratonovich representation and can not be obtained when
the basic Langevin equation is intended in the Ito sense. 

 In section VI we show evidence of a first order phase
transition at zero value of the spatial coupling.  This is, the
system, that in absence of spatial coupling is disordered,
develops a finite value of the order parameter even for 
infinitesimal values of the spatial coupling constant.
 For all the 
previously studied models the spatial
coupling has to be above a certain, non-zero, value to 
observe ordering. This 
supports the MN as the minimal model exhibiting a NIOT.

 Finally some conclusions are presented.

\section{Model definition and Connection with KPZ}

    In this section we define the multiplicative noise 
Langevin equation, and review some of its properties and 
connections with KPZ.
The MN equation is 
\begin{equation}
{\partial}_{t}\psi = - a \psi -
 p {\psi}^{p+1} + D {\nabla}^{2} \psi
+ \sigma \psi \eta 
\end{equation}
intended in the Stratonovich sense,
where $\psi(x,t)$ is a field, and $a$, $p$, $D$  and
$\sigma$ are parameters,
and
 $\eta$ a Gaussian white noise with
\begin{eqnarray}
\left< \eta (x,t) \right> &=& 0 \nonumber  \\
\left< \eta (x,t) \eta (x',t') \right> &=&
(1 - \alpha \psi^2) \delta (x-x') \delta (t-t').
\label{MN}
\end{eqnarray}
The Fokker-Planck equation associated with this reads 
\begin{eqnarray}
&&{d P(\psi(x),t) \over dt }  =  - \int dx {\delta \over \delta \psi(x)}
[- a \psi -
 p {\psi}^{p+1} + D {\nabla}^{2} \psi] \nonumber \\
&+ & {\sigma^2 \over 2} \int dx {\delta \over \delta \psi(x)}
\sqrt \psi (1 - \alpha \psi^2) {\delta \over \delta \psi(x)}
\sqrt \psi (1 - \alpha \psi^2) \nonumber \\
\end{eqnarray}

To simplify things, we could just consider the 
 $\alpha=0$ case for which we recover the 
{\it pure multiplicative noise} equation analyzed in \cite{MN,MN2}.
The equation with $\alpha>0$ was introduced in \cite{GMS}
as a prototype model exhibiting not only a NIOT
 but also a NIDT.
 This is, the order
parameter does not
keep on growing as noise amplitude is increased (as happens in the
case $\alpha=0$).
 Instead, it reaches
a maximum value after the NIOT, and decreases upon further 
increasing of noise amplitude, until a
NIDT transition appears and the systems comes
 back to a disordered state.
Phenomena of this type are often called 'reentrant transitions'.

 Although, in principle, we could work in the simplest case 
 $\alpha=0$, for technical reasons most of the
 numerical results present in what follows are
obtained for $\alpha=1$, but it is worth stressing that,
 apart from the presence of the 
NIDT, non of the (universal) results depend on $\alpha$.

  By performing a Cole-Hopf transformation ($n=\exp h$)
 this equation
(with $\alpha=0$)
reduces to:
\begin{equation}
{\partial}_{t} h(x,t) = - a - p  \exp( p h  ) + D {\nabla}^{2}h +
D ({\nabla} h)^2 +
  \eta. 
\label{KPZ}
\end{equation}
This is just a KPZ equation for a surface, defined
by the height variable $h(x,t)$,  except for the exponential term.
This acts as a wall 
repelling $h$ from positive to negative values \cite{dos}.
 For large values
of $a$ the surface escapes linearly in time from the wall
and therefore, asymptotically, any effect of it is lost and the
equation reduces to KPZ. In terms of $\psi$ the  unbounded
phase
corresponds to the
absorbing phase characterized by a vanishing value of 
its stationary order parameter value.
 On the other hand, for small enough
values of $a$ the surface remains bound to the wall (or wetting
the wall). In this the stationary value of $\psi$ takes
a non-vanishing value.  Separating both regimes there is a 
 critical point
which nature has been analyzed in \cite{MN,MN2}.  
Some exponents, unexisting for KPZ, can be defined for MN.
For example, if $m$ is the averaged order parameter,
$\xi$ the correlation length, and $\tau$ the correlation
time we have:
\begin{eqnarray}
\xi  &\sim & |a -a_c (\sigma)|^ {\nu_x} \\
\tau  &\sim&  |a -a_c  (\sigma)|^{\nu_t} \\
m (a) &\sim & |a -a_c  (\sigma)|^{\beta_a} \\
m(\sigma) &\sim& |\sigma -\sigma_c (a)|^{\beta_{\sigma}} \\
m(t,a=a_c, \sigma=\sigma_c)& \sim& t^{-\theta}.
\end{eqnarray}
 Some of these exponents can be related to KPZ exponents 
using scaling arguments \cite{MN,MN2}. For example, if 
 $z$ is the
dynamic exponent in KPZ,  
  it was proved in \cite{MN2}
that $\nu_x = 1/(2 z-2 )$ and $\beta_a > 1$. On the other hand,
using straightforward scaling relations we have $\theta=
\beta_a/\nu_t$ and $\nu_t = z \nu_x$.

Some other exponents can be defined in analogy with what is
customary in the study of systems with absorbing states \cite{Torre}. These
are the so called spreading or epidemic exponents. To measure
them one places an initial seed in an otherwise absorbing 
configuration and study the evolution of the space
 integral of $\psi$, $N(t)$, the surviving probability $P(t)$, and the 
mean square deviation from the origin $R^2(t)$. At the critical
point these scale as 
\begin{eqnarray}
N(t) & \sim & t^{\eta}\\
P(t)  & \sim & t^{-\delta}\\
R^2(t) & \sim & t^{z'}
\label{se}
\end{eqnarray}
where $\eta$, $\delta$ and $z'=2/z$
are the spreading exponents. The following scaling law is expected
 to hold \cite{delta,JFFM}:
\begin{equation}
\eta + \delta + \theta = d z'/2.
\label{HR}
\end{equation}
 Searching for power law behaviors of the spreading magnitudes is
a very precise way to determine the critical point.

  Given the aforementioned connections  between MN and KPZ, it is 
not surprising  that their respective renormalization group 
(RG) flow diagrams  
\cite{MN} resemble very much to each other.
In particular, for both of them \cite{Terry,MN}:
At any dimension larger than $d=2$ there are two different    
attractive fixed points: (i) a mean field or weak coupling (weak noise in
the MN language) fixed point 
at which the non-linear parameter vanishes, and (ii) a strong coupling 
(strong noise in
the MN language),
non-trivial fixed point, not accessible to standard perturbative techniques
\cite{Terry,Luc}. This means, that in particular, in $d=3$ there are
two different phases depending on the noise intensity: For small
intensities the system is in a weak coupling phase 
characterized by mean field like exponents. For noise
intensities above a certain critical threshold the system is in a 
(rough) strong coupling phase.   

   The multiplicative noise equation has been studied numerically
in $d=1$, and it was found that in fact the predicted relations
with KPZ hold. The best values for the different critical 
exponents are reported in table I.  However, as said before
 there is no roughening transition in $d=1$,
and consequently it has not been observed so far in systems 
with MN.   On what follows we  present the results of 
extensive numerical simulations performed in three dimensional
systems. By changing the noise amplitude we intend to observe
the two different phases: one with exponents related to the
strong coupling $3d$ KPZ exponents, and the other related 
to  mean field (i.e. Edward Wilkinson \cite{HZ,Laszlo}) exponents.

\section{Scaling analysis and mean field results}

  Let us start discussing in this section 
the mean field predictions for the previously defined 
exponents, which already present some interesting features,
and in the next section we will present numerical simulations
of the three-dimensional model.

Let us fist present some naive scaling arguments. For that 
we define  
the generating functional associated
 to Eq. (\ref{MN}) \cite{GF,ZJ}
\begin{equation}
Z=\int D\psi \ D\phi \exp{(- \int {d}^{d}{\bf x} \ dt \cal L)}
\label{FunzGen}
\end{equation}
with $\cal L$ given by
\begin{equation}
 {\cal L}  =  
  {\sigma^2 \over 2}{\phi}^{2}{\psi}^{2}  
 + \phi \left[{\partial}_{t}\psi+(a-{\sigma^2 \over 2}) \psi
+p{\psi}^{p+1}-{\nabla}^{2}\psi \right] .
\label{Lagrangiano}
\end{equation}
Using naive dimensional arguments,
the dimensions of the time $t$, $d_t$,
the field $\psi$, $d_{\psi}$, the
response field $\phi$, $d_{\phi}$ and $\sigma^2$,
$d_{\sigma^2}$ expressed in function of momenta 
(inverse of length) are
\begin{eqnarray}
{d}_{t}&=&-2 \nonumber \\
{d}_{\psi}+{d}_{\phi}&=&d \nonumber \\
{d}_{\sigma^2}+2d-d-2&=&0 \rightarrow {d}_{\sigma^2}=2-d.
\label{MNdim}
\end{eqnarray}
From this,  we conclude that the
noise amplitude is marginal at the critical 
dimension $d_c=2$, irrelevant above it and relevant below $d=2$, and
this result does not depend on the degree of the other nonlinearity,
i.e. on $p$.

As it was shown in \cite{delta}
the surviving probability in general systems with absorbing 
states scales as the response field. 
In the case of multiplicative noise the
particle density in  the absorbing state 
(i.e. in the unbound phase) decay continuously 
to zero, but 
does never reach that value
(in fact $h$ goes continuously
 to minus infinity, and
 for any finite though large value of $h$, $n$ takes a non zero value). 
Therefore 
the surviving probability is equal
 to unity and the dimension of the response
field is zero, $d_{\phi}=0$,
 (i.e. it scales as a constant)
\cite{delta,MN2}; this implies that $\delta=0$.
 Consequently $d_{\psi}=d$, which at the critical dimension 
is $d_{\psi}=2$; and therefore (using that the dimension of $a$ is
$2$)
\begin{equation}
m \sim {[a]}^{{\beta}_{a}}
 \rightarrow 2=2{\beta}_{a} \rightarrow {\beta}_{a}=1.
\label{betaauguno}
\end{equation}
 Analogously $\beta_{\sigma}=1$.
Observe that these results depend on the nature of the noise, and are
independent on the degree of the nonlinearity in Eq. (\ref{MN}), 
i.e. on $p$. This gives us a justification, of the fact, observed
numerically \cite{MN2} that Eq. (\ref{MN}) gives the same exponents
for different values of the nonlinearity $p$ 
 (this same property
is also shared by the exactly solvable zero dimensional case 
\cite{GS}).

   On the other hand, it is interesting 
to observe that naive mean field 
approximations, not coming from power counting of the 
generating functional 
 give the wrong result $\beta_a=\beta_{\sigma}=1/p$.
In particular, in appendix A, we present different types of 
mean field approaches all of them leading to the same (wrong)
prediction for the critical exponent $\beta$.   The origin for
the failure of standard mean field approaches is a rather delicate issue. 
We believe this is based on the fact, that even in the weak noise regime
 the stationary
probability distribution is non-trivial; in particular it is 
non-symmetric and its mean value is typically far away from the
most probable one. A detailed analysis of this and related issues
will be discussed elsewhere.

Regarding the rest of critical indices,
the mean field predictions are
as follows.
$\eta$ is the anomalous dimension
of the field, and therefore it vanishes in mean field
approximation where no diagrammatic corrections are
taken into account. 
For the same reason, just considering naive power counting 
arguments: $z'=1$, $z=2$, $\nu_x=1/2$, 
and $\theta= 1$.

 \section{Numerical Results: the roughening transition}

In this section we describe the results of extensive numerical simulations
of Eq.
(\ref{MN}) performed using the Heun method (see \cite{Max} and references 
therein). For that purpose space and time have been discretized
 using meshes of  ${a}_{r}=1$ (space) and $\epsilon =0.001$ (time)
respectively, and have fixed $p=2$.
In $d=1$ we have chosen $D=0.2$, and $a=1$ while in $d=2$
 and $d=3$ we take $D=1$,
and $a=1$ (weak noise regime) or $a=18$ (strong noise regime).
{\it In all dimensions we verify that the system exhibits
a NIOT as well as a NIDT for $\alpha >0$}.

\subsection{$d=1$}

We consider a system size  $L=1000$, $D=0.2$, $a=1$,
 the space and time meshes
are as said previously $1$ and $0.001$ respectively.

We determine some exponents,
the values
of which have not been previously reported in the literature,
in particular $\beta_{\sigma}$, and some other with improved precision,
 and illustrate the methods employed 
to compute them. 

In order to determine accurately the location of the critical point 
 ${\sigma}_{c}$ (keeping $a$ fixed and varying $\sigma$)
  and the critical exponent $\beta_{\sigma}$, we determine
numerically the order parameter as a function of $\sigma$ (see Fig.
\ref{fig1}). In order to measure the order parameter 
we let the system evolve long enough so the stationary state
is reached.   Then we write
$m={(\sigma-{\sigma}_{c})}^{{\beta}_{\sigma}}$, and take as critical point
the value of 
 ${\sigma}$ that maximizes the linear correlation coefficient  
when representing 
 $\log(m)$ as a function of $\log(\sigma -{\sigma}_{c})$ 
(see Fig. \ref{fig2} and
 \ref{fig3}).  From the corresponding slope we determine 
 ${\beta}_{\sigma}$ (Fig. \ref{fig3}). In particular we
obtain ${\sigma}_{c}=1.81 \pm 0.07$ and
 ${\beta}_{\sigma}=0.9 \pm 0.1$.

\begin{figure}
\protect
\centerline{\psfig{figure=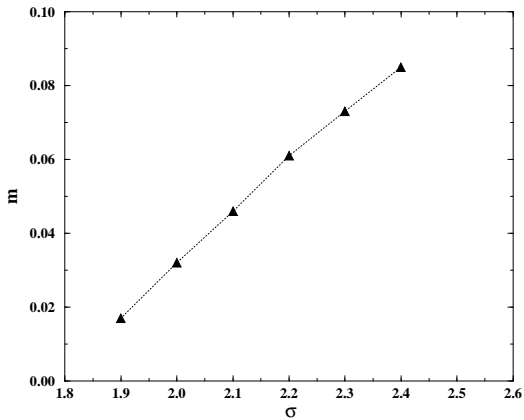,width=8cm}}
\seq
\caption{Order parameter $m$ as a function of $\sigma$
 in the vicinity of the critical point for $d=1$. 
}
\label{fig1}
\end{figure}

\begin{figure}
\centerline{\psfig{figure=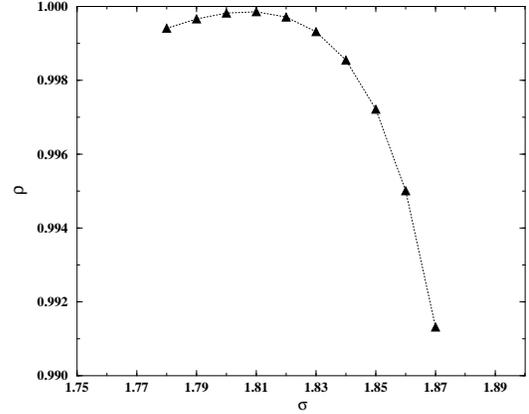,width=8cm}}
\seq
\caption{Linear correlation coefficient when fitting
 $\log(m)$ in $d=1$ as a function of $\log(\sigma -{\sigma}_{c}))$,
for different values of ${\sigma}_{c}$. The maximum of this curve gives  
the best estimation for the critical point
 $ {\sigma}_{c}=1.81 \pm 0.07$.
 }
\label{fig2}
\end{figure}

\begin{figure}
\centerline{\psfig{figure=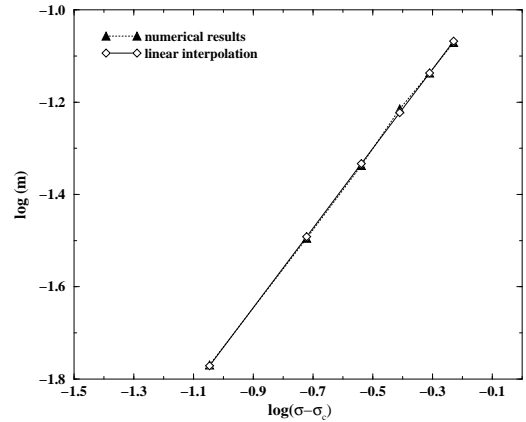,width=8cm}}
\seq
\caption{Order parameter $m$ as a function of $(\sigma-{\sigma}_{c})$
 (on a log-log scale)
in $d=1$. The slope gives the critical index ${\beta}_{\sigma}=
 0.9 \pm 0.1$.}
\label{fig3}
\end{figure}

In order to determine the exponent $\beta_a$, defined as
$m\propto {({a}_{c}-a)}^{{\beta}_{a}}$,
we fix $\sigma=\sigma_c$, and
diminish $a$ to stay in the active phase.
 Then we follow a maximization of the linear correlation 
coefficient procedure similar to the one described above.
 In that way we measure
 ${a}_{c}=1.04\pm 0.01$
 and  ${\beta}_{a}=1.5 \pm 0.1$.

Right at the critical point the order parameter decays in time as
$m(t)\propto t^{-\theta}$. In order to have an independent estimation of the critical
point we plot the
local slope of the averaged magnetization 
as a function of $1/t$ for different values of $\sigma$ (see Fig.(\ref{fig4})).
It is clear that the curve for $\sigma=1.8$
($\sigma=1.7$) curves
upward (downward) and corresponds to the active (absorbing) phase;
the critical point is located around $\sigma_c \approx 1.75$, slightly
smaller than the previously determined value,  but compatible
with that value within accuracy limits.
The intersection point at $1/t=0$ of the central curve gives
the value of the exponent $\theta$;  $\theta =1.1\pm 0.1$.

\begin{figure}
\centerline{\psfig{figure=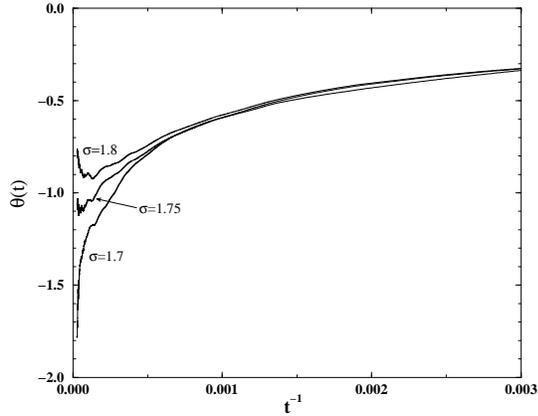,width=8cm}}
\seq
\caption{Local slope, $\theta (t)$, of $\log m(t)$ as a function of $\log t$,
plotted as a function of $t^{-1}$. The extrapolated value at $t^{-1}=0$
gives the value of $\theta$,  $\theta =1.1\pm 0.1$.}
\label{fig4}
\end{figure}

In order to measure spreading exponents we consider much larger system 
sizes.  Simulations are stopped either when 
the activity arrives to any of
 the system boundaries.
In this way we have determined 
 $z'$, $\eta$. 
Using the previously obtained valued $\sigma_c =1.75$, 
we measure $z'=1.25\pm 0.10$ e $\eta=-0.4\pm 0.1$. 
The standard critical exponent  $z$ is $z=\frac{2}{z'}=1.6\pm 0.1$ 
compatible with  the KPZ value $ z=1.5$ (in an analogous 
discrete model argued to be in the same MN universality class,
which is expected to converge faster to its asymptotic
behavior we measured $z=1.52 \pm0.03$ \cite{DNA}, which
remains
the most accurate estimate for $z$). For this system the exponent $\nu_x$
has been already measured numerically \cite{MN2,DNA}; the result
$\nu_x=1$ is in agreement with the theoretical prediction \cite{MN}.
  See Table I, for a
 complete list of the to the date best exponent estimates \cite{MN2,DNA,GMS}.
  Observe 
that all the scaling relations (including that for spreading
exponents) are satisfied within numerical accuracy.

\subsection{$d=2$}

  We have performed simulations in $d=2$ systems with $L^2=1600$,
and confirmed the presence of a phase transition (as it was already 
observed in \cite{GMS}), but have not performed extensive analysis
to determine accurately the critical exponents.
 At $d=2$ there are two fixed points of the RG for KPZ
a trivial, unstable, one at zero noise amplitude to which 
correspond,
obviously, mean field exponents, and a stable, rough phase one, 
with non trivial exponents for any non-vanishing noise amplitude.
 Instead of analyzing this case
with only one stable fixed point, 
we preferred to analyze the
 a priori more
interesting three-dimensional case.

\subsection{$d=3$}

In the three-dimensional simulations we consider system
sizes up to $L^3=64000$ and periodic boundary conditions.
 The space and time
 meshes are 
$1$ and $0.001$ respectively. The spatial coupling constant is
$D=1$.  

\subsubsection{The weak noise phase.} 

We fix $a=1$ for this small value we expect the transition to occur
at a small value of $\sigma$, and therefore to be controlled by the
weak noise fixed point (weak coupling, in he language of KPZ).
In the weak noise regime the mean field 
predictions (see appendix)
 are expected to be exact, and therefore for $a=1$ one should have
$1-\sigma^2/2=0$, implying $\sigma_c=\sqrt2$.
In fact, 
following the same procedure described in the one-dimensional case,
the best estimation of the critical point is  ${\sigma}_{c}=1.420\pm 0.002 $
(the deviation from $\sigma_c=\sqrt2$ is a finite size effect)
and the slope of a log-log plot of the order parameter versus $\sigma-
\sigma_c$ gives ${\beta}_{\sigma}=1.00\pm 0.01$  (see Fig. (\ref{fig5})).
\begin{figure}
\centerline{\psfig{figure=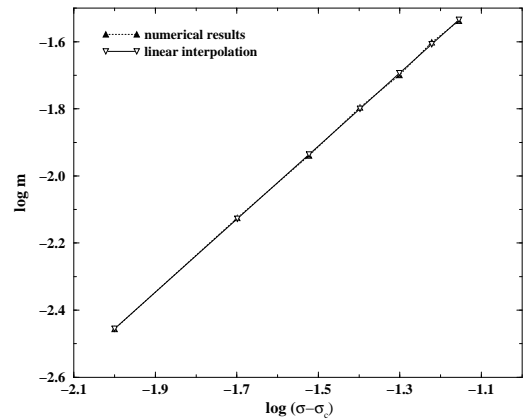,width=8cm}}
\seq
\caption{
Order parameter $m$ as a function of $(\sigma-{\sigma}_{c})$
 (on a log-log scale)
in $d=3$ with $a=1$ and $p=2$.
 The slope gives the critical index ${\beta}_{\sigma}=
1.00\pm 0.01$.}
\label{fig5}
\end{figure}
The order parameter time-decay exponent, $\theta$ is found to be 
 $\theta =1.0\pm 0.1$, while for $z'$ and $\eta$ we measure 
 $z'=1.00\pm 0.01$ (see Fig. ( \ref{fig6})) and $\eta= -0.1 \pm 0.1$.
Using scaling relations we estimate $\nu_x=0.50 \pm 0.05$.

\begin{figure}
\centerline{\psfig{figure=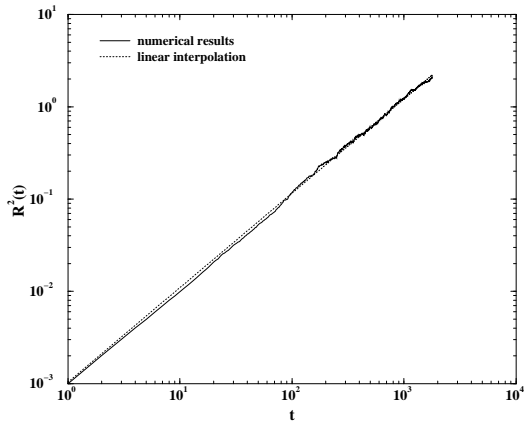,width=8cm}}
\seq
\caption{Log-log plot of $R^{2}(t)$ as a function of $t$
 at the critical point for $d=3$ and $a=1$. From the slope we obtain 
$z'=1.00\pm 0.01$.}
\label{fig6}
\end{figure}
 Fixing 
 $\sigma =1.418$ we have measured $m$
 for different values of $a$, with $a<1$. 
The linear correlation coefficient of
 a log-log plot of the order parameter versus
$a-a_c$ is maximum for $a_c =0.995 \pm 0.010$,
 and the corresponding slope 
gives ${\beta}_{a}=0.97\pm 0.05$, also compatible with its
mean field value ${\beta}_{a}=1$
(see Fig. (\ref{fig7})).

\begin{figure}
\centerline{\psfig{figure=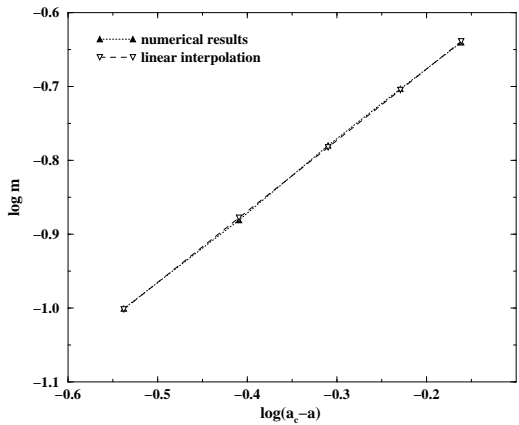,width=8cm}}
\seq 
\caption{$\log(m)$
 as a function of $\log({a}_{c}-a)$ and its corresponding linear
interpolation for $a=1$, $\sigma =1.418$ and $d=3$. From the slope 
we measure ${\beta}_{a}=0.97\pm 0.05$.}
\label{fig7}
\end{figure}
Therefore, summing up, {\it all the exponent in the weak noise
regime are
 in  good agreement with their corresponding
mean field values, and their expected scaling relations are satisfied.}

\subsubsection{The strong noise phase.}
 
 Now we take a large value of $a$, namely $a=18$, for which the transition
is expected to occur at a large value of the noise amplitude, and therefore 
to be controlled by a strong noise fixed point (strong coupling, 
in the KPZ language).
We find the critical point to be located at
 ${\sigma}_{c}=7.2\pm 0.3$, and ${\beta}_{\sigma}=1.2\pm 0.1$
 (see Fig. (\ref{fig8})). Observe that contrarily to the 
weak noise case, now the critical value of $\sigma$ is 
renormalized; the mean field prediction is $\sigma_c=\sqrt{2a}=6$.
\begin{figure}
\centerline{\psfig{figure=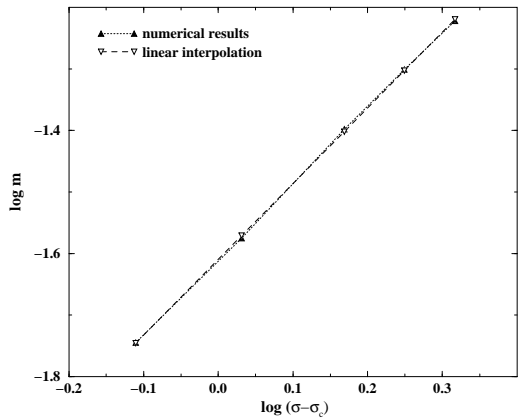,width=8cm}}
\seq
\caption{Log-log plot of the stationary value of the order parameter
$m$ as a function of $\sigma-\sigma_c$ for $a=18$ and $d=3$.
 From the slope
we measure ${\beta}_{\sigma}= 1.2\pm 0.1$ .}
\label{fig8}
\end{figure}
Following the previously described methods, we 
 we obtain $\theta =2.0\pm 0.1$, and for the spreading exponents
$z'= 1.20\pm 0.01$, and $\eta = -0.5 \pm 0.1$
 (the best power laws fit for the spreading exponents are obtained 
for $\sigma = 6.875$ slightly smaller that the critical value obtained 
from the order parameter analysis). Using the value of $z'$,
 we obtain
 $z=\frac{2}{z'}=1.67\pm 0.03$ in excellent agreement
with the best estimation 
for the strong noise phase of
 KPZ in $d=3$, namely $z=1.695$ \cite{Ala}. 
 And applying the scaling law
relating $z$ and $\nu_x$, we obtain.
 $\nu_x = 0.75\pm 0.03$.
 On the other hand the hyperscaling relation for spreading 
exponents Eq. (\ref{HR}) is not expected to hold above the upper critical 
dimension where dangerously irrelevant operators should affect it \cite{ZJ}
and, in fact, introducing the values obtained numerically one
observes that it is clearly violated.

 Fixing $\sigma$ to its critical value
 and varying $a$ we obtain ${\beta}_{a}=2.5\pm 0.1 $
(see Fig. 
 (\ref{fig9})); in particular, 
${\beta}_{a}\ge 1$ in agreement with the prediction made in \cite{MN}.
Contrarily to the mean field predictions observe that in the 
strong noise regime $\beta_{a} \neq \beta_{\sigma}$.

\begin{figure}
\centerline{\psfig{figure=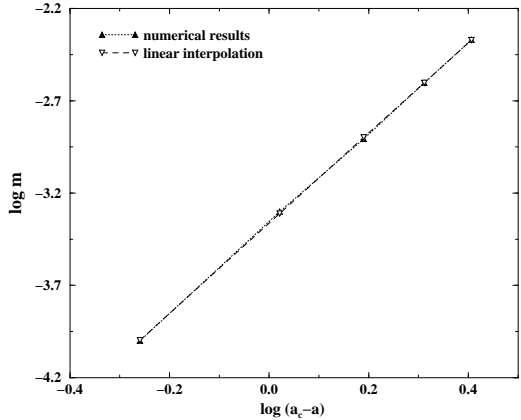,width=8cm}}
\seq
\caption{$\log(m)$ as a
 function of $\log({a}_{c}-a)$ and its corresponding linear 
interpolation for $\sigma =6.875$ and ${a}_{c}=18.05$, for $d=3$. 
From the slope we determine ${\beta}_{a}=2.5\pm 0.1 $.}
\label{fig9}
\end{figure}
Using 
$\nu_x = \beta_a /(z \theta)$ and introducing the
measured values of $\beta_a$, $\theta$, and $z$, we determine
$\nu_x = 0.76 \pm 0.03$ in excellent
agreement with our previous estimation.  This provides a test 
for the accuracy of our measurements. 

 In conclusion, {\it we have verified numerically the existence of two  
different regimes for the MN equation  exhibiting 
different values of the critical exponents, 
and
 related respectively to the 
weak and strong coupling regimes of the KPZ equation.}

\section{Ito-Stratonovich dilemma}

 In order to study the dependence of
the NIOT on the type of interpretation,
in the sense of the
Ito-Stratonovich dilemma, of the Langevin equation let us
now consider  Eq. (\ref{MN}) intended in the Ito sense.
It is obvious that an equation completely equivalent
to Eq.(\ref{MN}), i.e. with exactly the same physics, can
be written in the Ito interpretation using the well known
transformation rules \cite{VK,Gardiner}. The problem we study
here is different; we
analyze the same multiplicative noise
 Langevin equation in a different interpretation,
 i.e. Ito instead
of Stratonovich.

By repeating the mean field like approximations discussed 
in the appendix, but
using the Ito interpretation, one obtains the same final results
Eq. (\ref{MF1}), and  (\ref{MF2}) just by  substituting
$(\sigma^2/2 -a)$ by $-a$. Therefore for positive definite initial
conditions, and positive values of $a$ (i.e. values for which the
deterministic equation has $m=0$ as the only solution), there is
no non-trivial solution. This indicates that the {\it NIOT disappears
when intending the MN in the Ito sense}, and therefore, in the Stratonovich
interpretation it is due to the effective shift of the a-dependent term
in the stationary probability distribution when multiplicative noise
is introduced. In the same way, it is also straightforward to verify
by performing a linear stability analysis that the homogeneous 
solution $\psi=0$ is stable, contrarily to what happens in the
Stratonovich interpretation. The presence of an instability had
been identified as a key ingredient to generate 
noise induced transitions (see \cite{GMS} and references therein), and
therefore in absence of it no ordering is expected in the Ito 
interpretation.

We have verified this prediction
in numerical simulations. 

%In Fig. [\ref{fig10}] the dependence of the order parameter on the
%noise amplitude in a two-dimensional system is shown (the conclusion
%does not vary upon dimensionality changes).  The parameter have been fixed
%to $a=-1$, $D=10$ and $p=1$, which implies the presence of a constant
%non-vanishing solution of the associated deterministic equation. If instead
%we choose $a=1$ for which the deterministic solution is $m=0$, the system
%is observed to remain disorganized for all noise amplitudes. Therefore
%in the Ito interpretation the MN has only disorganizing effects.
%\begin{figure}
%\centerline{\psfig{figure=fig10.ps,width=8cm}}
%\seq
%\caption{Stationary value of the order parameter, $m$,
%as a function of $\sigma$ in a two-dimensional simulation of the MN
%Langevin equation in the Ito representation.}
%\label{fig10}
%\end{figure}

\section{ Coupling constant dependence}
 
  In this section we pose ourselves the question of which is the minimum
value of the coupling necessary 
to obtain a NIOT. As pointed out in \cite{Kawai,Raul}
the NIOT appears due to the interplay between a short time instability and
the presence of a spatial coupling 
that renders stable the generated non trivial
state.
 In all the previously discussed models exhibiting a NIOT and a NIDT there
are critical values of $D$ below which no ordering is possible. In order to 
determine whether there is a critical $D$ in our model we have studied 
it in $d=1$ by changing $D$ with fixed $a=1$ (the forthcoming results 
are qualitatively independent of the value of $a$).
 In Fig. (\ref{fig10}) we show a sketchy phase diagram,
outcome of systematic numerical simulations. 
\begin{figure}
\protect
\centerline{\psfig{figure=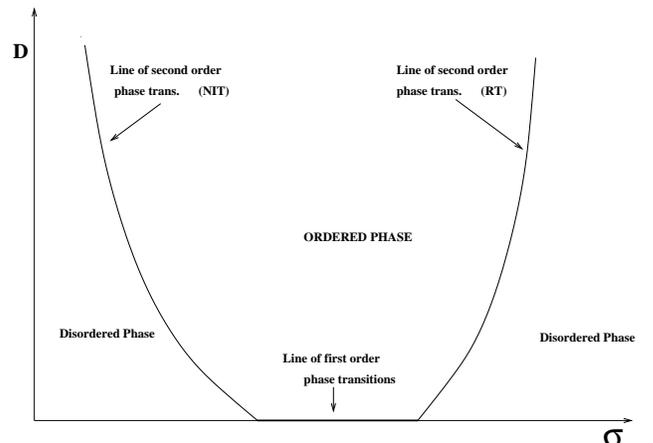,width=8.5cm}}
\seq
\caption{Schematic Phase diagram in the plane $(\sigma ,D)$ for the MN
Langevin equation. The rightmost line of second order phase transitions
moves to the right as $\alpha$ is reduced; and goes to $\infty$ in the
limit $\alpha=0$, indicating that the NIDT disappears. }
\label{fig10}
\end{figure}

There is a large interval of values 
of $\sigma$ for which the system exhibits a 
{\it first order} transition at $D=0$; 
i.e. as soon as an arbitrarily small spatial coupling is switched on the 
system gets ordered 
(for $D=0$ the only stationary state is $m=0$). 
In Fig.  (\ref{fig11})
the order parameter is plotted as a function
 of $1/D$ for a value of $\sigma$  in this
 interval; observe
how even for values as small as $ D=10^{-7}$, $m$ takes a large value
of about  $0.15$.
\begin{figure}
\protect
\centerline{\psfig{figure=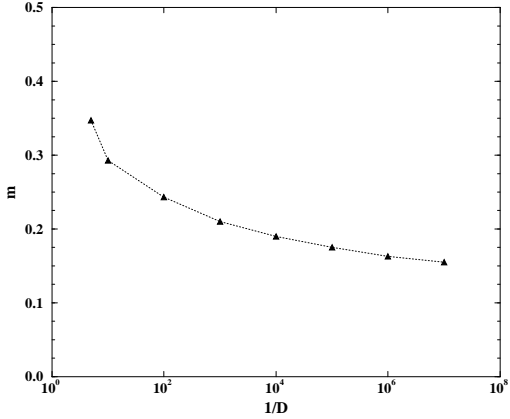,width=8cm}}
\seq
\caption{Order parameter $m$ as a function
 of $1/D$ for $\sigma =10$ and $a=1$. The curve converges to a constant for
large values of $1/D$, this is for small couplings, indicating that
at $D=0$ there is a first order phase transition.}
\label{fig11}
\end{figure}

  The fact that there is a large interval for which the systems
becomes ordered as soon as a spatial coupling is switched on,
property that was absent in all the previously studied models 
for NIOT, is a new indication that the MN equation is the minimal 
model for NIOT, and that the associated ordering mechanism is not
mixed up with other unnecessary ingredients.

  For values of $\sigma$ out of
 the previously discussed interval, the system exhibits  
a  {\it second order} phase transitions at a  value of $D$, $D_c(\sigma)$,
 Defining $\beta_D$ by
\begin{equation}
m\propto {\left(D-{D}_{c}\right)}^{{\beta}_{D}},
\label{centocinquantasei}
\end{equation}
we obtain 
 ${D}_{c}=0.094\pm 0.003$ and
  ${\beta}_{D}=0.08\pm 0.03$ for the particular
choice of parameters $a=\sigma=5$.

\section{Conclusions}

  We have presented some recent results on Langevin equations 
with multiplicative noise. In particular, we have studied
numerically for the first time the presence of two 
different regimes:  Weak and a strong noise regimes, 
in $d=3$. All the predicted scaling laws and relations with KPZ exponents 
in its respective weak coupling and strong coupling fixed points 
are verified.  On he other hand, we have shown that the noise
induced ordering transition associated with Langevin equations
with multiplicative noise is specific of the Stratonovich representation,
and shown that these noise
induced ordering transitions are obtained even for arbitrarily small
values of the spatial coupling constant, supporting the fact
that the Langevin equation with pure multiplicative noise is
the minimal model for  noise
induced ordering transitions.

\section{Appendix}

In this appendix we present some different mean field approximations
to evaluate the order parameter exponent for both $\alpha =0$ and   
$\alpha >0$.

Defining the averaged magnetization as $m$,
in the limit of large dimensionalities the discretized Laplacian operator
can be written as
\begin{equation}
  \nabla^2 \psi = 1/2d \sum_{j, N.N.} \psi_j - \psi_i
\approx m -\psi_i.
\end{equation}
 Using this approximation, and determining $m$
in a self-consistent way it is possible to obtain an analytical
solution of the Langevin equation.
 In what follows we present 
different calculations corresponding to infinite and finite
values of the spatial coupling $D$ respectively. In both cases
the obtained value of the critical exponent $\beta_a$ is 
$1/p$. 

\subsection{Infinite spatial coupling limit: $D \rightarrow \infty $}

   Using the previous approximation, writing down the stationary
probability distribution, solution
of the associated Fokker-Planck equation \cite{VK,Gardiner},
 and imposing  the
 self-consistent requirement $m = <\psi>$ we obtain 

\begin{equation}
\displaystyle{
 m=\frac{
\int_{I} d\psi \psi
 \exp
 \int_{0}^{\psi} d\psi
 \frac{F+D(m-\psi ) - GG'/2}{G^2/2} 
 }        
 { \int_{I} d\psi
\exp  \int_{0}^{\psi} d\psi \frac{F+D(m-\psi )-GG'/2}{G^2/2}
        }}.
\label{centododici}
\end{equation}
For large values of $D$ the integral can be evaluated in
saddle point approximation \cite{Raul,Kawai}, giving
\begin{equation}
m={\left[ \frac{1}{p} \left(
\frac{{\sigma}^{2}}{2}-a \right) \right]}^{\frac{1}{p}}.
\label{MF1}
\end{equation}
 The NIOT transition is predicted at $\frac{{\sigma}^{2}}{2}=a$
 with an associated exponent
${\beta}_{a} = {\beta}_{\sigma}=\frac{1}{p}$.

\subsection{Finite spatial coupling}

In order to make sure that the previous result is not due to 
the approximation involved in considering $D \rightarrow \infty $,
we present here an analogous calculation for finite values of $D$.
In this case, the associated asymptotic stationary probability is
\begin{equation}
P_{\infty}\propto {\psi}^{-1-\frac{2}{{\sigma}^{2}}(a+D)}
\exp{\left(- \frac{2 D m}
{{\sigma}^{2}\psi}\right) }
\exp{\left(- \frac{2}{{\sigma}^{2}}{\psi}^{p}\right)}
\label{MNProbas}
\end{equation}
where $m$  has to be fixed self-consistently
 by imposing $m=\langle \psi \rangle
 $, this is
\begin{equation}
\displaystyle{m=\frac{
\int_{0}^{\infty}d\psi 
{\psi}^{-\frac{2(a+D)}{{\sigma}^{2}}}
\exp\left(-\frac{2Dm}{{\sigma}^{2}\psi}
-\frac{2}{{\sigma}^{2}}{\psi}^{p}\right)}{\int_{0}^{\infty}
d\psi {\psi}^{-1-
\frac{2(a+D)}{{\sigma}^{2}}}
\exp\left(-\frac{2Dm}{{\sigma}^{2}\psi}-
\frac{2}{{\sigma}^{2}}{\psi}^{p}\right)}.}
\label{MNautocons}
\end{equation}
The numerical solution of this last equation for parameter values
 $a=1, D=1$, and $ p=1$ is shown in Fig.
 \ref{fig12} and Fig. 
\ref{fig13}. 

\begin{figure}
\centerline{\psfig{figure=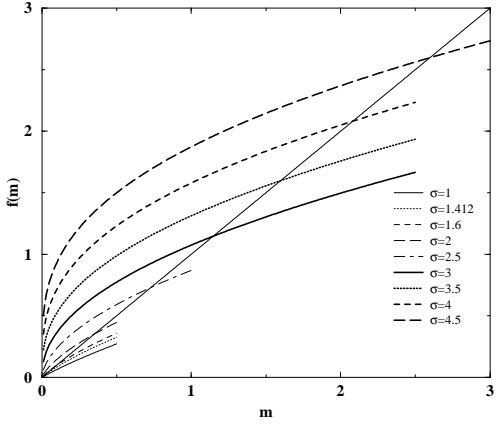,width=8cm}}
%\seq
\caption{The solution, $m$, of the self-consistency equation is the
intersection point between
 $y=f(m)$ (where $f(m)$ represents the function
on the r.h.s. of Eq. ()) and $y=m$.}
\label{fig12}
\end{figure}

\begin{figure}
\centerline{\psfig{figure=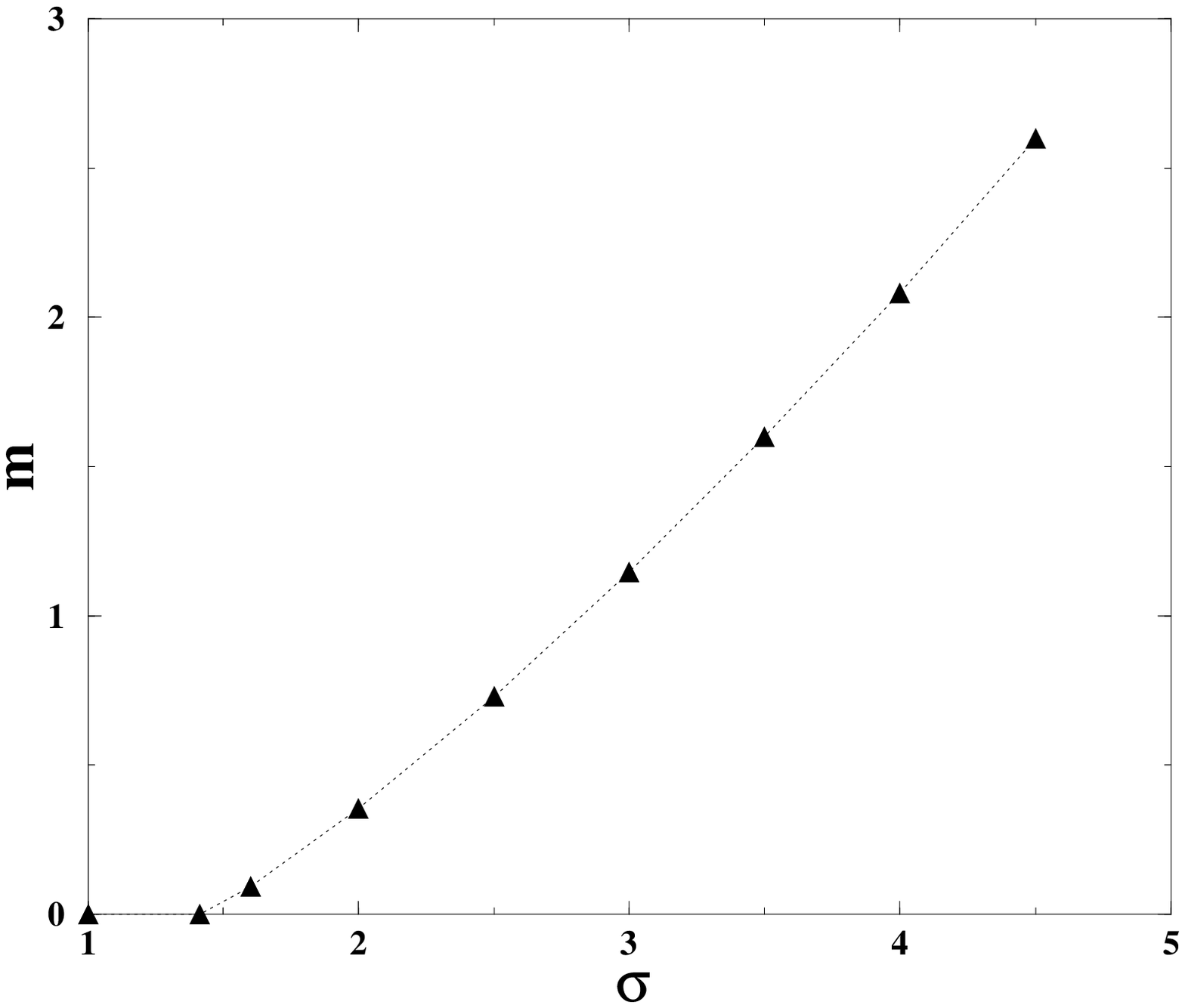,width=8cm}}
%\seq
\caption{
$m$ as a function of $\sigma$
in mean field theory with
$\alpha=0$.
The points here correspond to the intersection of
 the curves
in the previous figure
with the line $m=f(m)$. 
The critical point is located
at $\sigma_c^2=2$. }
\label{fig13}
\end{figure}

In order to derive the exponent $\beta_a$ in this MF approximation we
define the following change of variables,
\begin{eqnarray}
t&=&\frac{1}{\psi}\nonumber \\
\alpha &=& \frac{2}{{\sigma}^{2}} \nonumber \\
\gamma &=& \frac{2}{{\sigma}^{2}}(a+D)  \nonumber \\
\mu &=& \frac{2Dm}{{\sigma}^{2}}.
\label{MNdef}
\end{eqnarray}
 Eq. (\ref{MNautocons}) can be written as
\begin{equation}
\displaystyle{\frac{\mu}{\alpha D}=
-1/{\partial}_{\mu}\log{ \left\{ \int_{0}^{\infty}dt \
 {t}^{\gamma -2}\exp \left[-\mu t-\frac{\alpha}{{t}^{p}}\right]\right\}}.}
\label{MNautomod}
\end{equation}
Introducing a Gaussian
 transformation the integral in the previous expression can
be rewritten as
\begin{eqnarray}
&&\int_{0}^{\infty}dt
 \ {t}^{\gamma -2+\frac{p}{2}}\exp[-\mu t]\int_{-\infty}^{+\infty}
d\eta \ \exp\left[-\frac{{t}^{p}{\eta}^{2}}{4\alpha}+i \eta\right]= \nonumber \\
&& \ \ \ \ \ \ 
 ={\mu}^{1-\gamma}\int_{-\infty}^{+\infty}d\eta \
 \exp\left(i \eta \sqrt{4\alpha {\mu}^{p}} \right) \nonumber \\
&&\ \ \ \ \ \ \ \ \ \ \ \
 \ \ \ \ \ \ \ \ \
 \int_{0}^{\infty}\frac{dt}{t} \ {t}^{\gamma +\frac{p}{2} -1}
\exp\{-t-{t}^{p}{\eta}^{2} \}= \nonumber \\
&&\ \ \ \ \ \ ={\mu}^{1-\gamma}
\int_{0}^{\infty} d\eta 
 \cos\left(\eta \sqrt{4\alpha {\mu}^{p}}\right) \nonumber \\
&&\ \ \ \ \ \ \ \ \ \
 \ \ \ \ \ \ \ \ \
 \ \int_{0}^{\infty}\frac{dt}{t} \ {t}^{\gamma +\frac{p}{2} -1}
\exp\{-t-{t}^{p}{\eta}^{2} \}=  \nonumber \\
&&={\mu}^{1-\gamma}\{ {H_{0}}^{(p)}(\gamma +\frac{p}{2}-1)
-2\alpha {\mu}^{p}{H_{1}}
^{(p)}(\gamma +\frac{p}{2}-1)+...\} \nonumber \\
&&
\label{MNgaussint}
\end{eqnarray}
where we have expanded the cosine function up to second order,
 and we have defined
\begin{eqnarray}
{H_{n}}^{(p)}(\delta )
&=&\int_{0}^{\infty} \ {\eta}^{2n}{\Gamma}_{p}(\delta ,{\eta}^{2}) \\
{\Gamma}_{p}(\delta ,{\eta}^{2})&=&
\int_{0}^{\infty}\frac{dt}{t}{t}^{\delta}\exp{\{-t-{t}^{p}{\eta}^{2}\}} 
\label{MNgammadef}
\end{eqnarray}
for $\delta > 0$.
  This calculation is valid only if
 $\gamma +\frac{p}{2}>1$.
 At the end of the calculation we will verify that  this constraint is
verified.
Eq. (\ref{MNautomod}) can be simply expressed as
\begin{equation}
 \alpha D = 
 2\alpha p {{H_{1}}^{(p)}(\gamma + \frac{p}{2}-1)
\over {H_{0}}^{(p)}(\gamma +\frac{p}{2}-1){\mu}^{p} }+(\gamma -1).
\label{MNautofinale}
\end{equation}
From this we find the solution
$\mu =0$ corresponding to
  $m=0$ and if $  \frac{{\sigma}^{2}}{2}-a \ge 0$ a second
solution exists with
\begin{equation}
m=\frac{{\sigma}^{2}}{2D}{\left( \frac{{H_{0}}^{(p)}}{2{H_{1}}^{(p)}
}\right)}^{\frac{1}{p}}{\left[\frac{1}{p}
\left(\frac{{\sigma}^{2}}{2}-a\right)\right]}^{\frac{1}{p}}
\propto {\left( \frac{{\sigma}^{2}}{2}-a \right)}^{\frac{1}{p}}.
\label{MF2}
\end{equation}
This solution confirms the results obtained in the
 $D\rightarrow \infty$ case; namely
 ${a}_{c}=\frac{{\sigma}^{2}}{2}$,
for $a\ge 0$, ${{\sigma}_{c}}^{2}=2a$,
and
${\beta}_{\sigma}={\beta}_{a}=\frac{1}{p}$ which is consistent with the
requirement $\gamma + p/2 > 0$ for all the values of $p$ rendering consistent
the calculation.

\subsection{Infinite coupling limit for $\alpha > 0$ }

For completeness' sake let us present here the mean field analysis
in the case in which $\alpha>0$. In this subsection we evaluate the
infinite coupling limit.

  The solution $m=0$ is unstable for
 ${\sigma}^{2}>2a$, and the new stable solution is
\begin{equation}
m={\left[\frac{{\sigma}^{2}-2a}{(p)
({\sigma}^{2}+1)}\right]}^{\frac{1}{p}}.
\label{MN-RT-Dinf}
\end{equation}
Let us notice that this approximation predicts a NIOT at the same
point the pure MN equation
(with $\alpha=0$)  does, namely $\sigma^2=a/2$, but contrarily to
the pure model the order parameter does not grow indefinitely
by increasing noise amplitude. Instead it saturates to 
a value $m=(p)^{-1/p}$.

\subsection{ Finite coupling for $\alpha > 0$}

In the case of finite
coupling $D$ and $\alpha >0$ we have that 
the asymptotic probability defined in the interval
 $0\le \psi \le 1$ is (for $p=1$)
\begin{eqnarray}
&&P_{\infty}(\psi)\propto
 {\psi}^{-\left(1+\frac{2(a+D)}{{\sigma}^{2}}\right)}
{\left(\frac{1-\psi}{1+\psi}
\right)}^{-\frac{Dm}{{\sigma}^{2}}} \nonumber \\
&&\ \ \ \ \ \
 \ \ \ \ \ \ \ \ \ \
{(1-{\psi}^{2})}^{-\frac{1}{2}+\frac{a+D+1}{{\sigma}^{2}}}
\exp\left[-\frac{2Dm}{{\sigma}^{2}\psi}\right].
\label{MNRTprobas}
\end{eqnarray}
The self-consistency equation is obtained equating $m$ to
\begin{equation}
\displaystyle{\frac{\int_{0}^{1}d\psi
\ {\psi}^{-\frac{2(a+D)}{{\sigma}^{2}}}
{(\frac{1-\psi}{1+\psi})}^{-\frac{Dm}
{{\sigma}^{2}}}{(1-{\psi}^{2})}^{-\frac{1}{2}+\frac{a+D+1}
{{\sigma}^{2}}}e^{-\frac{2Dm}{{\sigma}^{2}\psi}}
}{\int_{0}^{1}{\psi}^{-(1+\frac{2(a+D)}
{{\sigma}^{2}})}{(\frac{1-\psi}{1+\psi})}^{-\frac{Dm}{{\sigma}^{2}}}
{(1-{\psi}^{2})}^{-\frac{1}{2}+\frac{a+D+1}
{{\sigma}^{2}}}e^{-\frac{2Dm}{{\sigma}^{2}\psi}}
}
.}
\label{MNRTauto}
\end{equation}
Both of the integrals exhibit a singularity at
 $\psi =1$, but they are integrable.
It is straightforward verifying that
in the limit $\sigma \rightarrow \infty$, $m \rightarrow 0$, and
therefore for finite values of $D$
this approximation predicts both
a NIOT (also located at $\sigma^2 =2 a$) and
a NIDT.
\begin{figure}
\protect
\centerline{\psfig{figure=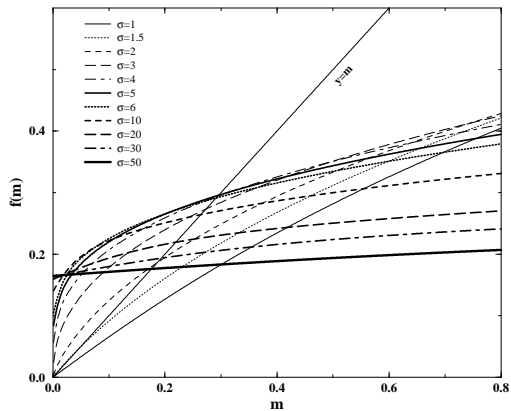,width=8cm}}
\seq
\caption{Solution of the mean field theory for $\alpha > 0$.
Observe that contrarily to what happens
in Fig. 13, here the intersection
point between the curves for different
 values of $\sigma$ and the straight
line $y=m$ reaches a maximum
 value after which it starts decreasing.}
\label{fig14}
\end{figure}
\begin{figure}
\protect
\centerline{\psfig{figure=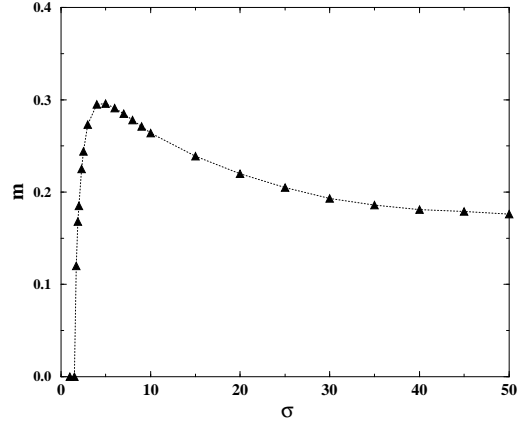,width=8cm}}
%\seq
\caption{ 
$m$ as a function of $\sigma$ in mean field theory with
$\alpha > 0$. The points here correspond to the intersection of the curves
in the previous figure with the line $m=f(m)$.
 The critical point is located at $\sigma_c^2=2$.}
\label{fig15}
\end{figure}

\vspace{1cm}
{\bf \ \ \ ACKNOWLEDGEMENTS}  
\vspace{1cm}

We acknowledge useful discussions with
P. Garrido, G. Grinstein, Y. Tu, T. Hwa, 
J. M. Sancho, L. Pietronero, R. Dickman, G. Parisi 
and R. Toral. We thank 
Claudio Castellano and R. Pastor-Satorras for a critical 
reading of the manuscript.
 M.A.M. is supported by the M. Curie fellowship contract
ERBFMBICT960925, and by the TMR 'Fractals' network,
project number EMRXCT980183.

\begin{table}
\centering
\begin{tabular}{|c|c|c|c|}
\hline
\rm d & $1$ & $3  \ (Weak N.)$ & $3  \ (Strong N.)$ \\
\hline
\hline
${\beta}_{a}$ & $1.5 \pm 0.1$ & $0.97 \pm 0.05$ & $2.5 \pm 0.1$  \\
\hline
${\beta}_{\sigma}$ & $0.9 \pm 0.1$ & $1.0 \pm 0.01$ & $1.2 \pm 0.1$ \\
\hline
$z$ & $1.52 \pm 0.03$ & $2.00 \pm 0.05$ & $1.67 \pm 0.03$ \\
\hline
$\eta $ & $-0.4 \pm 0.1$ & $-0.1 \pm 0.05$ & $ -0.5 \pm 0.1$ \\
\hline
$\theta $ & $1.1 \pm 0.1$ & $1.0 \pm 0.1$ & $ 2.0 \pm 0.1$ \\
\hline
$\nu_x $ & $1.0 \pm 0.1$ & $0.50 \pm 0.05$ & $ 0.75\pm 0.03$ \\
\hline
\end{tabular}
\caption{Table of critical indices obtained from numerical simulations. 
In the second column
we report results for the $d=1$ case.
 In the  three-dimensional weak noise phase (third column)
the exponents are in very good agreement
 with the expected mean field values.
The last column reports the
 numerical values obtained in his work for the 
three-dimensional exponents in the strong noise phase.}
\end{table}
\end{text}

\begin{references}
\bibitem{dos} In the case $\alpha>0$ there are two different walls:
the one discussed in the text, and another one at a smaller value,
at $ {\bar h}= -{1/2}\ln(\alpha)$.
As $\alpha$ goes to $0$ this second wall goes to $- \infty$. 
By performing a change of variables $\Upsilon = (1 + \sqrt{\alpha} \psi)$,
and $\bar{\eta} = \eta/\sqrt{\Upsilon (2-\Upsilon)}$, the Langevin
equation for $\Upsilon$, has a noise amplitude proportional to
$\sqrt{\Upsilon (2-\Upsilon)}(\Upsilon -1) \propto \sqrt{\Upsilon}$. 
Even if the noise amplitude is proportional to the
square root of the field,
the critical phenomena associated at this NIDT are not
expected to be related to the Reggeon field theory (RFT)
 and directed percolation
\cite{RFT}. This is because this second wall, in spite of having a
RFT like noise is not truly absorbing, i.e. the dynamics does
not cease even for a flat configuration at ${\bar h}$.
\end{references}
\end{document}